\documentclass[journal]{IEEEtran}
\usepackage[pdftex]{graphicx}
\usepackage{amssymb,amsfonts,amsmath,verbatim,url,multirow}
\usepackage{pbox}
\begin{document}

\title{Predicting Financial Markets: Comparing Survey, News, Twitter and Search Engine Data}

\author{ Huina~Mao,~\IEEEmembership{Indiana University-Bloomington,}~Scott~Counts,~\IEEEmembership{Microsoft Research,} and Johan~Bollen,~\IEEEmembership{Indiana University-Bloomington}}

\maketitle

\begin{abstract}
Financial market prediction on the basis of online sentiment tracking has drawn a lot of attention recently. However, most results in this emerging domain rely on a unique, particular combination of data sets and sentiment tracking tools. This makes it difficult to disambiguate measurement and instrument effects from factors that are actually involved in the apparent relation between online sentiment and market values. In this paper, we survey a range of online data sets (Twitter feeds, news headlines, and volumes of Google search queries) and sentiment tracking methods (Twitter Investor Sentiment, Negative News Sentiment and Tweet \& Google Search volumes of financial terms), and compare their value for financial prediction of market indices such as the Dow Jones Industrial Average, trading volumes, and market volatility (VIX), as well as gold prices. We also compare the predictive power of traditional investor sentiment survey data, i.e. Investor Intelligence and Daily Sentiment Index, against those of the mentioned set of online sentiment indicators. Our results show that traditional surveys of Investor Intelligence are lagging indicators of the financial markets. However, weekly Google Insight Search volumes on financial search queries do have predictive value. An indicator of Twitter Investor Sentiment and the frequency of occurrence of financial terms on Twitter in the previous 1-2 days are also found to be very statistically significant predictors of daily market log return. Survey sentiment indicators are however found not to be statistically significant predictors of financial market values, once we control for all other mood indicators as well as the VIX. 
\end{abstract}


\section{Introduction}

\IEEEPARstart{T}{he} \emph{efficient market hypothesis} (EMH) asserts that financial market valuations incorporate all existing, new, and even hidden information, since investors act as rational agents who seek to maximize profits. Behavioral finance \cite{kahneman-tversky-1979} has challenged this notion by emphasizing the important role of behavioral and emotional factors, including \emph{social mood} \cite{Prechter99}, in financial decision-making. As a consequence, measuring investor and social mood has become a key research issue in financial prediction. 

Traditionally, public and investor mood are measured by surveys. For example, the Gallup Life Evaluation Index measures the general well-being of the US public on a daily basis by conducting a survey across a representative sample of the US population. Investor mood is likewise assessed by surveys, in which investors or newsletter writers rate their current stance on the market, e.g. Daily Investor Sentiment \footnote{\url{www.trade-futures.com}} and Investor Intelligence \footnote{\url{http://www.investorsintelligence.com/x/us_advisors_sentiment.html}}. In spite of their popularity, surveys are, however, resource intensive and thus expensive to conduct, and can be subject to problems related to responder truthfulness \cite{Da2010,singer2002}, individual biases, social biases, and group think.

In recent years, researchers have explored a variety of methods to compute indicators of the public's sentiment and mood state from large-scale online data. This approach holds considerable promise. First, computational analysis of public sentiment or mood may be more rapid, accurate and cost-effective to conduct than large-scale surveys. Second, there now exists considerable support for the claim that the resulting public mood and sentiment indicators are indeed valid measurements of public sentiment and mood, even to the degree that they have been found to predict a variety of socio-economic phenomena, including presidential elections \cite{Tumasjan2010}, commercial sales \cite{Choi2009,Mishne2006}, and influenza epidemics \cite{Culotta:epidemics2010}. It is of considerable interest to behavioral finance that a respectable and growing amount of literature in this area has shown that computational indicators of public sentiment may also have predictive value with respect to financial market movements \cite{Antweiler2004,BollenMao2011,Bordino2011:websearch,Da2010,Da2011,Preis2010}. 

To the best of our knowledge, three distinct classes of online data sources have been investigated for financial prediction. First, \emph{news media} content has been shown to be an important factor shaping investor sentiment. For instance, Tetlock found that high levels of pessimism in the Wall Street precede lower market returns the following day \cite{Tetlock2007}. This effect has also been observed at the level of individual firms, with high negative sentiment forecasting lower firm earnings \cite{Tetock08:language}. In \cite{Schumaker09:textmining} it was shown that adding textual features of news to a stock prediction system can improve the forecasting accuracy.

Second, \emph{web search (query) data} has been shown to be related to and even predictive of market fluctuations. Search volumes of stock names reveal investor attention and interest, and high search volumes thus predict higher stock prices in the short-term, and price reversals in the long-term \cite{Da2010}. Also, search volumes of stocks correlate highly with trading volumes of the corresponding stocks, with peaks of search volume anticipating peaks of trading volume by one day or more \cite{Bordino2011:websearch}. Similar phenomena have been found at the weekly level \cite{Preis2010}.  

Third, \emph{social media feeds} are becoming an important source of data to support the measurement of investor and social mood extraction. In an early study, Internet stock message boards were studied to predict market volatility and trading volumes \cite{Antweiler2004}. In past couple of years, public mood indicators extracted from social networks such as Facebook \cite{Karabulut2011}, LiveJournal \cite{Gilbert2010} and Twitter \cite{BollenMao2011} have been used to predict stock market fluctuations. 

Together these results are highly suggestive that a variety of web-scale data sources may provide predictive power in financial analytics. However, each of the mentioned investigations uses different types of web data to predict different financial indicators. It is not clear which mood indicators constructed from particular data sources most effectively capture investor mood-related signals and thereby provide the best predictive power.

In this paper, we therefore collect multiple data sources, i.e. surveys, news headlines, search engine data and Twitter feeds, from which we define a variety of sentiment indicators, i.e. Survey Investor Sentiment, Negative News Sentiment, Google search volumes of financial terms, Twitter Investor Sentiment and Tweet volumes of financial terms. Subsequently, we determine the predictive value of these sentiment indicators over a range of financial indicators, i.e. Dow Jones Industrial Average price, trading volumes, market volatility (VIX) and the price of gold.

\section{Data Collection and Sentiment Analysis}
In this section we outline our data collection methods, and how we computed investor sentiment indicators from Twitter, news, and search engine data. 

\subsection{Survey Data}
Surveys are the most direct and common method for collecting investor sentiment. Investor Intelligence (II), published by an investment services company, determines whether opinion in over one hundred independent market newsletters points towards a bullish, bearish or correction market. II has been available at a weekly level dating to 1964. Daily Sentiment Index (DSI) provides daily market sentiment readings on all active US markets daily since 1987, and is one of the most popular short-term market sentiment indices for futures traders. High vs. low DSI values of respectively above 90\% or below 10\%, suggests that a short-term top or bottom is either developing, or has been achieved. 

\subsection{News Media}
We chose eight news media outlets to collect our news data from: Wall Street Journal, Bloomberg, Forbes.com, Reuters Business\& Finance, BusinessWeek, Financial Times, CNNMoney and CNBC. These are the top news sources for financial traders and investors. In order to track recent and featured news from these sources, we followed their respective Twitter accounts (\emph{``wsjusnews"}, \emph{``wsjbreakingnews"}, \emph{``wsjmarkets"}, \emph{``bloombergnews"}, \emph{``bloombergnow"}, \emph{``bloomberg"}, \emph{``forbes"}, \emph{``BusinessWeek"}, \emph{``Reuters Business"}, \emph{``reuters\_biz"}, \emph{``financialtimes"}, \emph{``FinancialTimes"}, \emph{``CNNMoney"}, \emph{``CNBC"}). We then extracted and parsed the URLs from these tweets, saving the story headlines as our news corpus. This approach of using headlines is based on previous research that studied stock price reaction to news headlines \cite{chan2003:newsheadlines}. 

\begin{figure}
	\begin{center}
		\includegraphics[width=9cm,height=6cm]{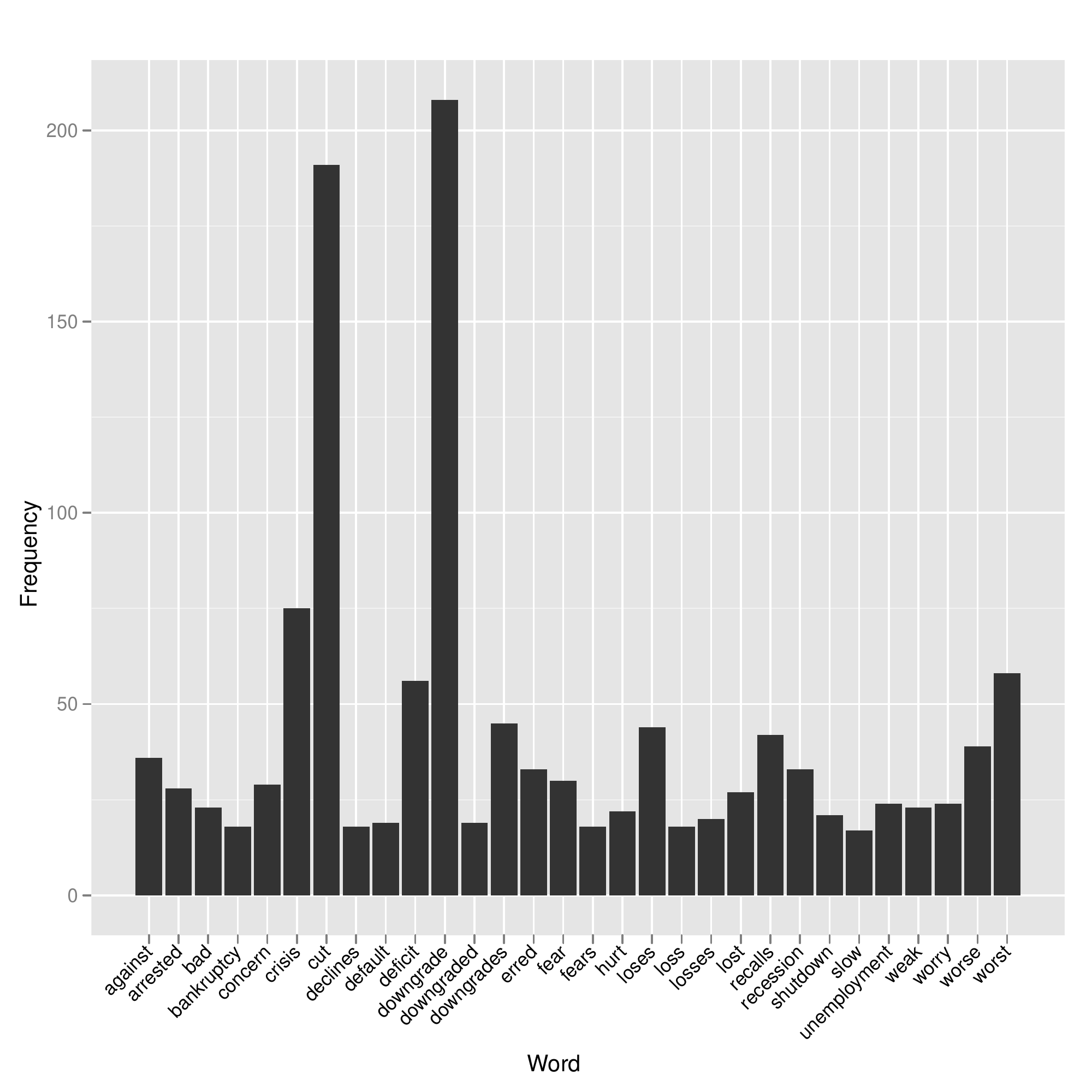} 
		\caption{\label{wf} Frequency of negative terms in News headlines from July 31st to August 9th 2011.}
	\end{center}
\end{figure}

Previous research has demonstrated that negative mood seems to be more predictive of financial market values than positive mood \cite{Tetlock2007}. There are two well-accepted financial lexicons for negative word identification. One is the Harvard IV-4 dictionary \footnote{\url{http://www.wjh.harvard.edu/~inquirer/}} as used in \cite{Tetlock2007, Tetock08:language}. The other \footnote{\url{http://www.nd.edu/~mcdonald/Word_Lists.html}} is developed by Loughran and McDonald in \cite{Loughran2011:liability}, which is shown to better reflect the tone of financial text than the Harvard IV-dictionary. In our paper, we apply the latter financial negative lexicon to our news headlines. We count the total number of words in a news headline and take the ratio of the number of \emph{negative sentiment} words to the total number of words in the headline. Then, we sum the emotional ratio and divide by the total number of news articles on the same day, yielding our \emph{Negative News Sentiment} score.  Fig. \ref{wf} shows the example of top negative financial terms of the news headlines from July 31st to August 9th 2011, when the DJIA dropped while market volatility increased. As a result words such as \emph{``downgrade"}, \emph{``cut"},\emph{``crisis"} and \emph{``losses"} frequently occur in news headlines in that period. 

\subsection{Search Engine Data}
\label{sec:searchengine}
Previous research has shown that search volume itself can be a mood indicator for financial market \cite{Bordino2011:websearch,Da2010,Da2011,Preis2010}. In \cite{Da2010}, it has been shown that the more people search on economic negative terms such as \emph{``recession"} and \emph{``bankruptcy"}, the more pessimistic people feel about the economy. To create a search query-based indicator of financial mood, we took the following steps. First, we downloaded the weekly search volume data for a set of seed queries including \emph{``dow jones"}, \emph{``stock market"}, \emph{``stock to buy"}, \emph{``stock"}, \emph{``bullish"}, \emph{``bearish"}, \emph{``financial news"} and \emph{``wall street"} from Google Insights for Search (GIS)\footnote{\url{http://www.google.com/insights/search/}}. GIS is a Google service that provides search volume data from January 2004 to the present. Second, to more fully capture search activity related to the financial markets we expanded these seed keywords with those terms that are top relevant search terms as recommend by GIS. This procedure resulted in a lexicon of about 26 financial search terms  for which we again retrieved GIS search frequency indices, resulting in a time series of GIS frequencies for all searches containing those 26 terms as shown in Table \ref{st}. 

\begin{table}[!htb]
		\caption{26 Search Terms\label{st} }
		\begin{center}
		\begin{tabular}{c}
		\hline
DJIA, Dow,Dow Jones, Dow Jones Industrial Average,
\\
bearish, bear market,best stock, bullish, bull market, 
\\
finance, finance news, financial news, financial market, \\
long stock, SP500, stock, stock market, stock decline, stock fall,
\\
stock market crash, stock market news, stock market today, \\
stock price, stock to buy, wall street, wall street news today \\


\hline
	\end{tabular}
	\end{center}
	\end{table}

\subsection{Social Media Data}
The enormous amount of social media data that has become available in recent years has provided significant research opportunities for social scientists and computer scientists. In fact, Twitter, which is now one of the most popular microblogging services, has been extensively used for real time sentiment tracking and public mood modeling \cite{bollen10:mood,Golder11:nature}. And its financial predictive power has also been explored. In \cite{Huberman2010} , it has been shown that Twitter content and sentiment can be used to forecast box-office revenues of movies. In \cite{zhang:coins2010}, the correlation between emotional tweets and financial market indicators are studied, indicating that the percentage of emotional tweets is significantly negatively correlated with Dow Jones, NASDAQ and S\&P500 values, but positively correlated with VIX values. Moreover, in \cite{BollenMao2011} a six-dimensional model of public emotions is derived from Twitter  (Calm, Alert, Sure, Vital, Kind and Happy) and found to have significant predictive power with respect to DJIA fluctuations.

In this paper, we use a 15\%-30\% random sample of all public tweets posted every day from July 2010 to September 2011. From this collection, we define two Twitter-based financial mood indicators: Twitter Investor Sentiment (TIS) and Tweet volumes of financial search terms (TV-FST). These are discussed in greater detail below.

\subsubsection{Twitter Investor Sentiment}
We simply define a tweet as bullish if it contains the term \emph{``bullish"}, and bearish if it contains the \emph{``bearish"}. On the basis of the number of Bearish and Bullish tweets on a given day, we define the investor sentiment score, Twitter Investor Sentiment ($TIS$) on day $t$, denoted $TIS_{t}$ as follows:

\begin{equation}
TIS_{t}=\frac{N_{bull}}{N_{bull}+N_{bear}}
\label{tis}
\end{equation}

where $N_{bull}$ is the number of bullish tweets on day $t$ and $N_{bear}$ is the number of bearish tweets on day $t$ .

\subsubsection{Tweet Volumes of Financial Search Terms (TV-FST)} 
\label{sec:tv-fst}
\begin{figure}[!htb]
	\begin{center}
		\includegraphics[width=9cm,height=7cm]{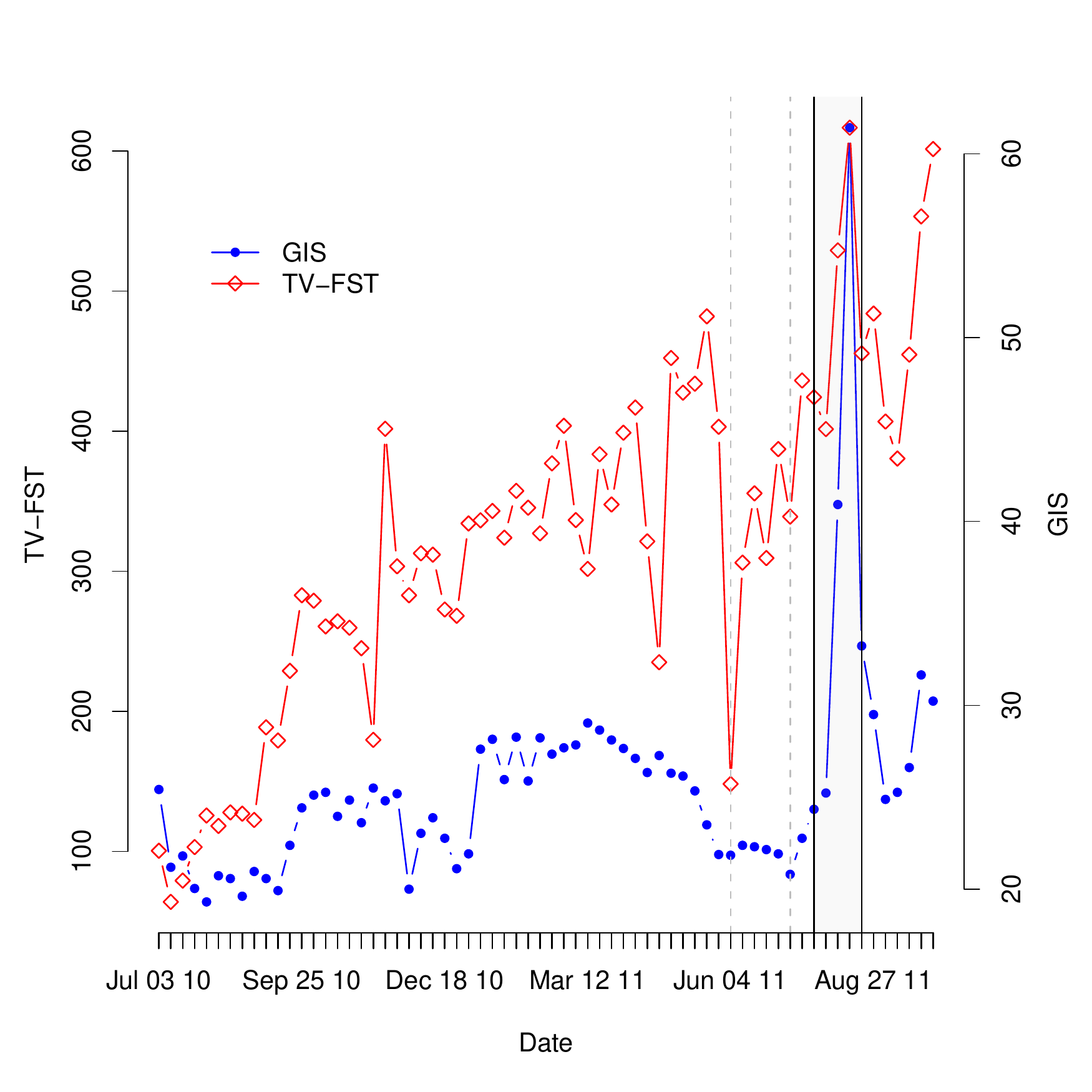}          
		\caption{\label{TwittervsGIS}Weekly TV-FST vs. GIS.}
	\end{center}
\end{figure}

As mentioned in Section~\ref{sec:searchengine}, search query volume of stock names and various financial/economic terms has been used in previous research as proxies of public and investor mood. Our proposal is to apply a similar approach to define our Tweet Volumes of Financial Search Terms indicator (TV-FST). We want to compare Tweet volumes and Search volumes of the same search queries. To do so, we use the following procedure for data processing: First, we compute both the weekly Google search volumes (GIS) and daily Tweet volumes of those 26 financial search terms from July 2010 to September 2011. Second, we calculate the weekly mean over the daily volumes of tweets. This step is necessary to compare Twitter (daily) and GIS (weekly) at the same time scale. Third, and finally, we take the average of the separate, weekly time series generated for each individual term, which yields a GIS and Tweet volume time series over 66 weeks, for the combination of all the financial search terms. Fig. \ref{TwittervsGIS} shows these two time series. 

A correlation analysis over all weekly values of the two time series reveals a statistically significant Pearson correlation coefficient of 0.62 ($p<0.01$). To see whether these two indicators signal notable movement in the financial market, we marked the time period from July 23rd to August 20th 2011 in a shaded rectangle as shown in Fig. \ref{TwittervsGIS}. During this period, the stock market had a huge decline (i.e., the DJIA declined 1864 points between July 22nd and August 19th 2011.) We can see that from June 4th, 2011 (at the first vertical line), TV-FST values started to increase, while 5 weeks later, on July 9th 2011 (at the second vertical line), GIS followed. This suggests that GIS may be less efficient than Twitter in revealing public/investor negative sentiment.

\subsection{Economic and Financial Market Data}
We collected daily and weekly Dow Jones Industrial Average, trading volume, Volatility (VIX) from Yahoo! Finance. In addition, we calculate the market log returns $R$ of stock prices $S(t)$ over a time interval $\Delta t$ as follows:
\begin{equation}
R_{\Delta t}=logS(t+\Delta t) - logS(t)
\label{logreturn}
\end{equation}                 
  
Here $\Delta t= 1$. Additionally, we also retrieved the price of gold \footnote{\url{http://www.gold.org/investment/statistics/gold_price_chart/}} over the same period of time. Table \ref{datanote} summarizes the corresponding time range and daily/weekly scale for all the data we obtained. 

\begin{table}[!htb]
	\begin{center}
		\caption{\label{datanote}Time-range coverage of different data sources.}
		\begin{tabular}{ccc}
		\hline
		Data     		&  Daily(mm/dd/yy) 	&    Weekly(mm/yy)					 \\\hline
		DSI (Daily Sentiment Index)         & 07/01/10 -09/05/11 		&  /			\\ \hline
		II (Investor Intelligence)        	&   /  	& 01/08 - 09/11
\\ \hline
		TIS (Twitter Investor Sentiment)          &  07/01/10 – 09/29/11  		& /
\\ \hline
\pbox{20cm}{TV-FST (Tweet volumes \\of financial search terms)} & 07/01/10 – 09/29/11 & /
\\ 
\hline
		NNS(Negative News Sentiment)          & 07/01/10 – 09/29/11  	& /  			
\\ \hline
		GIS (Google Insight Search)       & /    	& 01/08 - 09/11 
\\ \hline
DJIA/VIX/Volume/Gold & 07/01/10 – 09/29/11 & 01/08 - 09/11 
		\\\hline
	
		\end{tabular}
	\end{center}
	\end{table}

\section{Search Volume (GIS)-based prediction of financial indicators}
\label{sec:weeklyAnalysis}
\subsection{Search Volume and Financial Indicator Correlations}
In this section, we compare the GIS time series (search query volume of 26 financial search terms) with the DJIA price, volume, and the price of gold from January 2008 to September 2011, roughly 196 weeks. This period was punctuated by significant market volatility, as well as significant bear and bull markets, thus allowing us to perform our analysis under a variety of market conditions.

We first compute the pair-wise correlation between our 26 time series of GIS search terms and the financial time series. All time series are transformed to log scale for analysis. The results are summarized in Table \ref{topGIS}. Due to the space limitations, we only list the correlations of 10 search terms. 

\begin{table}[!htb]
	\begin{center}
		\caption{\label{topGIS}Pearson correlation coefficients between GIS and VIX, DJIA, Trading Volume.}
		\begin{tabular}{cccc}
		\hline
		Search Query     		& VIX   		&   DJIA     	&  Volume				 \\\hline
		\emph{DJIA}          & 0.88    		&  -0.76        	& 0.69    					\\
		\emph{Dow Jones}       	&   0.84   	& -0.69    		&   0.68     						\\
		\emph{Dow}        &  0.83   		& -0.67    		& 0.68 		
\\
		\emph{Dow Jones Industrial Average}           & 0.78     	& -0.77   		& 0.65   		 			\\
		\emph{Stock market news}      &  0.77     	& -0.37   		& 0.59  						\\
		\emph{Finance}        & 0.71    	& -0.50   		& 0.70   		\\
\emph{Stock market today} & 0.69 & -0.62 & 0.51 \\
\emph{Financial news} & 0.68 & -0.43 & 0.57 \\
\emph{Stock} & 0.66 & -0.38 & 0.57 \\
\emph{SP500} & 0.65 & -0.34 & 0.49 

\\\hline
		\end{tabular}
	\end{center}
	\end{table}

We find relatively strong correlations in most cases, especially for what seem to be DJIA-relevant search terms such as \emph{``DJIA"}, \emph{``Dow Jones"}, etc. The GIS time series has a positive correlation with the VIX and trading volumes, but negative correlations with DJIA, which may indicate that as more people search on financial terms, the market will be more volatile (i.e. high VIX), and trading volumes will be higher, while DJIA prices will move lower. 
 
\begin{figure}[!htb]
	\begin{center}
		\includegraphics[width=9cm,height=10cm]{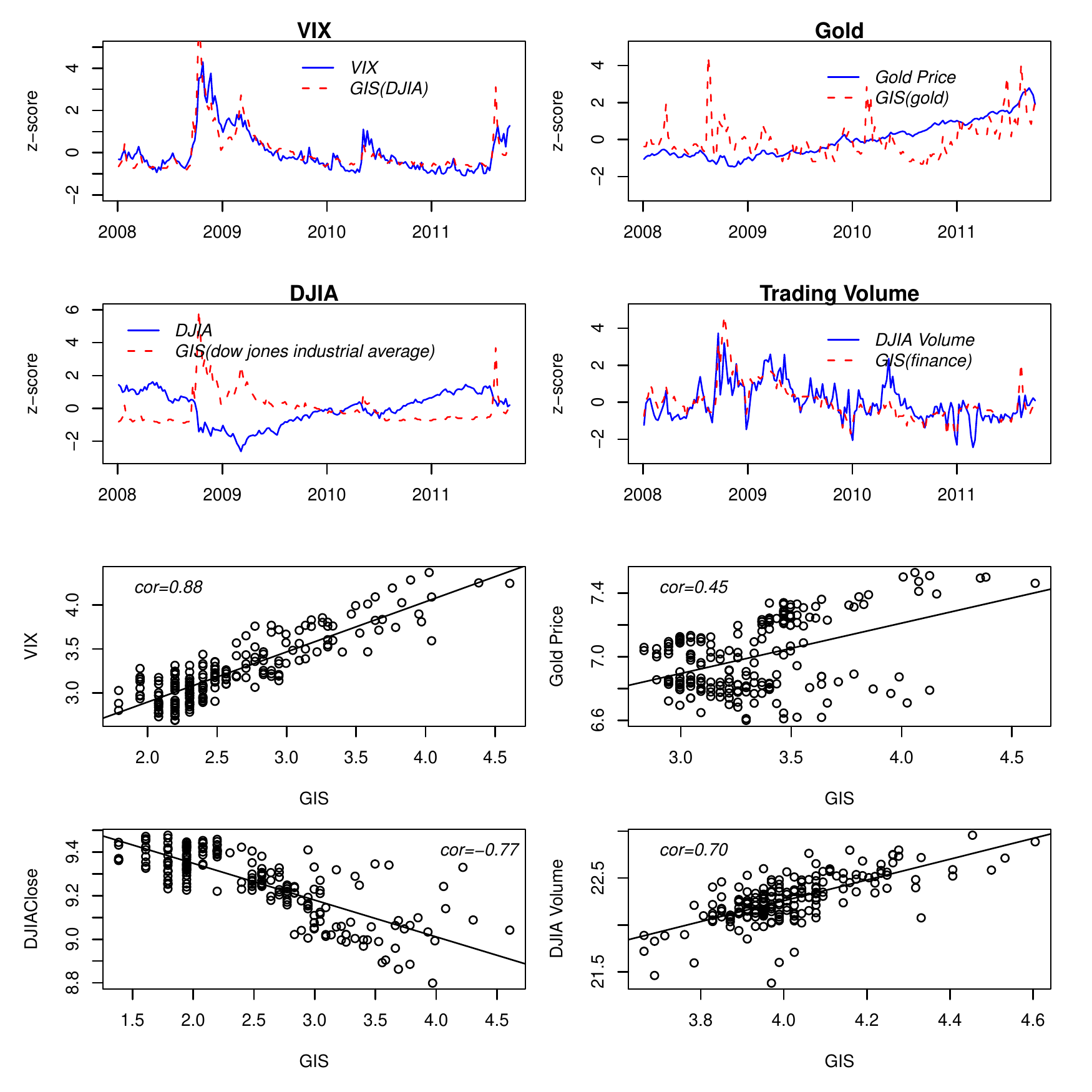} 
		\caption{\label{scatterplot}Trend analysis and log scale scatter plots of GIS time series vs. financial indicators such as VIX, DJIA closing values, gold price and DJIA trading volume. (Search query terms are inside the brackets).}
	\end{center}
\end{figure}

For further testing, we keep the top search term whose search volume has the highest correlation with the corresponding financial index for each time series. In Fig. \ref{scatterplot}, we overlaid the resulting time series with the mentioned financial indicators to visually examine the occurrence of any particular trend. 

The top panels of Fig. \ref{scatterplot} show the actual time series whereas the lower panels show the scatter plot of GIS values vs. financial indicator values in log-log scale. A simple visual inspection of the top panels reveal a clear correlation between GIS search term volumes and the financial indicator time series; peaks in GIS values generally co-occur with those of VIX and Volume values, and in some cases even precede the peaks of the various financial time series (DJIA, Gold). The scatter plots in Fig. \ref{scatterplot} show that search volumes exhibit a high positive correlation with VIX and trading volume ($\gamma=0.88$, $\gamma=0.70$), and a high negative correlation with DJIA price ($\gamma=-0.77$). The correlation between gold price and search volumes on \emph{``gold"} is also satisfactory ($\gamma=0.45$). This correlation value may in fact be an underestimation due to non-linear patterns in how the two variables relate. For $log$(gold prices) $>7.0$ we do observe a linear pattern of correlation. Below that value there seems to be little to no correlation. This pattern is confirmed by the trend plot at the upper right of Fig. \ref{scatterplot}: from mid-2010 to the end, at higher gold prices, we indeed observe a strong positive correlation, and in fact two spikes of search volumes appear before the gold price reached its peak in early September 2011. 

VIX is a widely used measure of market risk and is often referred to as the ``investor fear gauge". Our results show that search volumes of financial terms reflect VIX fluctuations, implying that search volume for key financial terms may be a computational gauge of ``investor fear".

To evaluate time-lag correlations between search volume and financial time series, we compute their cross-correlation.  In order to compare the effectiveness of search volumes with the survey data with respect to how well they predict the financial markets, we also include the Investor Intelligence (II) time series in our analysis. 

Consider two series $x=\{x_{1},...,x_{n}\}$ and $y=\{y_{1},...,y_{n}\}$, the cross correlation $\gamma$ at lag $k$ is then defined as:
\begin{equation}
\gamma=\frac{\sum_{i}(x_{i+k})-\bar{x})(y_{i}-\bar{y})}{\sqrt{\sum_{i}(x_{i+k})-\bar{x})^2}\sqrt{\sum_{i}(y_{i}-\bar{y})^2}}
\label{corr}
\end{equation}

where $\bar{x}$ and $\bar{y}$ are the sample mean values of the $x$ and $y$, respectively. We use the cross-correlation function provided in $ccf$, an R statistics package. For example, where   $ccf(x,y)$ estimates the correlation between $x[t+k]$ and $y[t]$, it means that we keep $y$ still, but move $x$ forward or backward in time by a lag of $k$. Where $k>0$, it means $y$ anticipates $x$, and vice versa. 

\begin{figure}[!htb]
	\begin{center}
		\includegraphics[width=9cm,height=6cm]{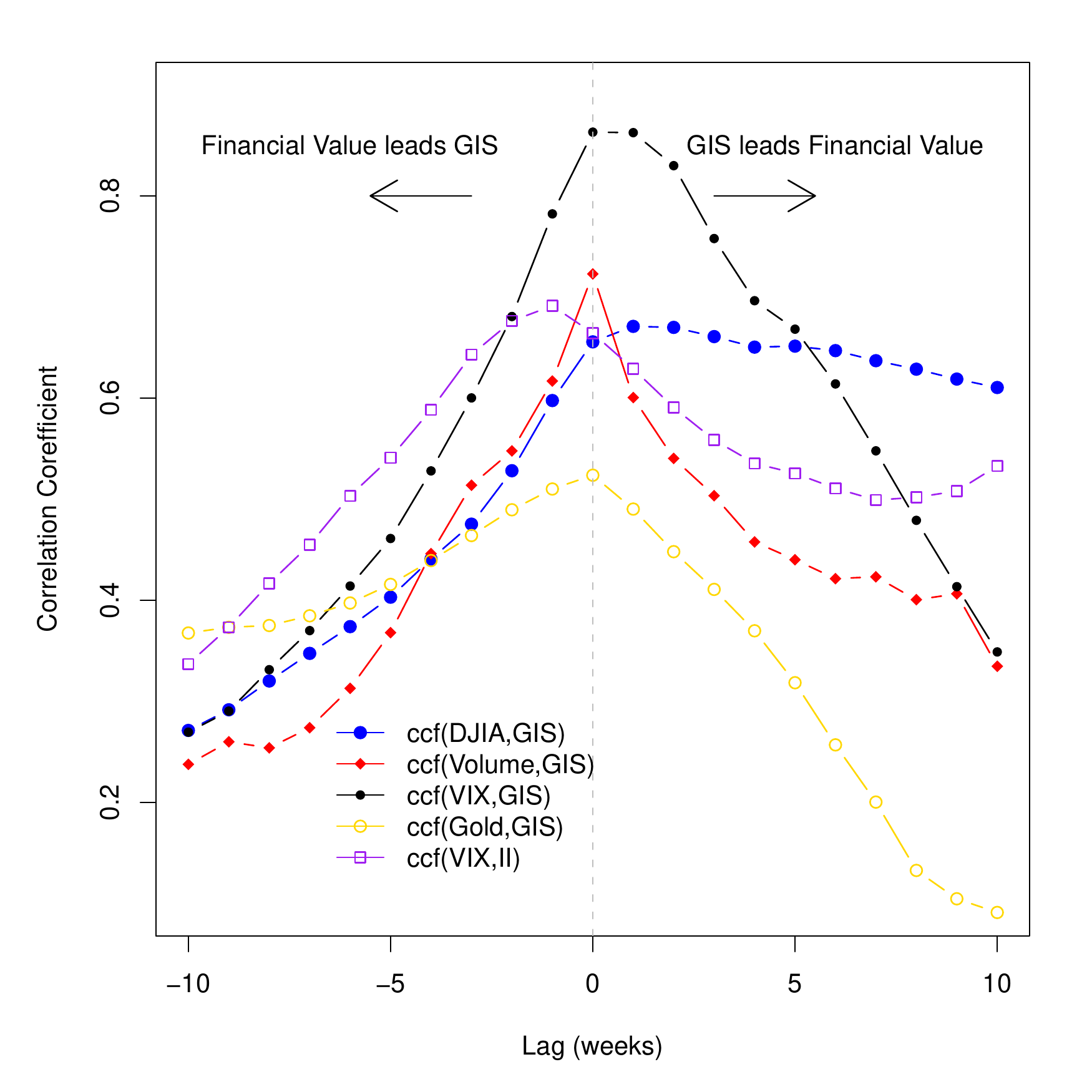} 
		\caption{\label{ccfFinGIS} Cross correlation analysis between financial time series and search volume (GIS) time series.}
	\end{center}
\end{figure}

As can be seen in Fig. \ref{ccfFinGIS}, DJIA values and GIS (search volume) exhibit the highest correlation and particularly so on the right side of the graph where lag values are positive, i.e. $k>0$, and, in other words, GIS values lead DJIA values. A similar effect can be observed for GIS vs. VIX values, especially where $k=[+1,+3]$ weeks. In contrast, as shown in Fig. \ref{ccfFinGIS}, the cross correlation between II and VIX seems to work in the opposite direction, indicating that VIX leads changes in II values. The correlation coefficients at both sides seem to be roughly balanced for trading volume. The search query time series for \emph{``gold"} exhibits the opposite effect of other search query time series: GIS search volumes on \emph{``gold"} do not lead gold prices. This runs counter to our earlier observation (in Fig. \ref{scatterplot}) that spikes of \emph{``gold"} search volumes precede spikes in gold prices, indicating that \emph{``gold"} GIS may yet have predictive value under certain conditions. We speculate this may be due to a non-linear interaction with absolute gold price levels, but we leave this for future exploration. 

\subsection{Granger Causality Analysis}

We further refine the observations discussed above by a Granger causality test, a technique that is widely used to analyze the relations between economic time series. The Granger causality test is a statistical hypothesis test to determine whether a time series $X(t)$ is useful in forecasting another time series $Y(t)$ by attempting to reject the null hypothesis that $X(t)$ does not help predict, i.e. Granger-cause, $Y(t)$. The alternative hypothesis is that adding $X(t)$ does help predict $Y(t)$. An F-test is conducted to examine if the null hypothesis can be rejected.  

We caution that Granger causality analysis might establish that the lagged value of $X(t)$ exhibits a statistically significant correlation with $Y(t)$. However, correlation does not prove causation. In other words, Granger causality testing does not establish actual \emph{causality}, merely a statistical pattern of lagged correlation. This is similar to the observation that cloud cover may precede rain and may thus be used to predict rain, but does not itself actually cause rain.

Table \ref{grangerpGIS} presents the results of applying the Granger causality test in two directions, i.e. with positive and negative lags, reflecting the hypothesis that each time series may Granger cause the other. 

\begin{table}[!htb]
	\begin{center}
		\caption{\label{grangerpGIS}Statistical significance (p-values) of Granger causality analysis between search volumes/ II and financial indicators over lags of 1, 2, and 3 weeks.}
		\begin{tabular}{cccc}
		\hline
		    		& 1   		&   2     	&  3				 \\\hline
		VIX$\rightarrow$GIS          & 0.0051$\star\star\star$    		& 0.0004$\star\star\star$        	& 0.0010$\star\star\star$
    					\\
		GIS$\rightarrow$VIX      	&   0.0025$\star\star\star$   	& 0.0202$\star\star$    		&   0.0091$\star\star\star$       						\\
		  VIX$\rightarrow$II      &  8.04e-05$\star\star\star$    		& 3.63e-07$\star\star\star$      		& 9.98e-08$\star\star\star$  		
\\
		II$\rightarrow$VIX           & 0.398     	& 0.726   		& 0.849  		 			\\
		DJIA$\rightarrow$GIS     &  0.207     	& 0.040$\star\star$   		& 0.096$\star$  						\\
		GIS$\rightarrow$DJIA        & 7.85e-04$\star\star\star$     	& 1.48e-03$\star\star\star$    		& 9.31e-04$\star\star\star$   		\\
Volume$\rightarrow$GIS & 0.409 & 0.705 & 0.843 \\
GIS$\rightarrow$Volume& 0.020$\star\star$ & 0.028$\star\star$  & 0.101 \\
Gold$\rightarrow$GIS & 0.055$\star$ & 0.104 & 0.082$\star$ \\
GIS$\rightarrow$Gold & 0.139 & 0.00036$\star\star\star$  & 0.0013 $\star\star\star$

\\\hline
\multicolumn{4}{l}{ ($p-value < 0.01$: $\star\star\star$, $p-value< 0.05$: $\star\star$, $p-value< 0.1$: $\star$)}
		\end{tabular}
	\end{center}
	\end{table}

The values in the first column of Table \ref{grangerpGIS} represent the particular hypothesis under consideration. For example, ``VIX$\rightarrow$GIS" represents the null hypothesis that adding VIX does not help predict GIS. As can be seen from the listed $p$-values, this particular null-hypothesis is rejected with a high level of confidence. In the row below, we observe that adding GIS can also help predict VIX. However, the Granger causality between Investor Intelligence (II) and VIX runs in only one direction, i.e. VIX$\rightarrow$II: adding survey data (II) does not help predict VIX. In addition, the null hypothesis that adding GIS does not help predict DJIA, is strongly rejected at a high level of confidence level. Similarly, we find a very significant $p$-value for GIS$\rightarrow$Gold at lag 2 and 3 weeks. GIS of the previous 1 to 2 weeks significantly Granger-cause trading volume.  

\subsection{Forecasting Analysis}
\label{sec:weeklyPred}
Can search volumes predict future values of financial indicators? As a further validation, we conduct a 1-step ahead prediction over 20 weeks based on a baseline model, denoted $M_{0}$, and an advanced model, denoted $M_{1}$. Here $Y$ represents the particular financial index (i.e. DJIA, trading volumes or VIX) and $X$ represents a sentiment indicator. In this section we will focus on GIS in particular. 

\begin{equation}
M_{0}: Y_t=\alpha+\sum_{i=1}^{n}\beta_{i}Y_{t-i}+\epsilon_{t}
\label{M0}
\end{equation}   

\begin{equation}
M_{1}: Y_t=\alpha+\sum_{i=1}^{n}\beta_{i}Y_{t-i}+\sum_{i=1}^{n}\gamma_{i}X_{t-i}+\epsilon_{t}
\label{M1}
\end{equation}   

Forecasting accuracy is measured in terms of the Mean Absolute Percentage Error (MAPE) and the direction accuracy. The MAPE is defined as follows:

\begin{equation}
MAPE=\frac{\sum_{i}^{n}|\frac{y_{i}-\hat{y_{i}}}{y_{i}}|}  {n}\times{100}
\label{MAPE}
\end{equation}   

where $\hat{y_{i}}$ is the predicted value and $y_{i}$ is the actual value. 
Direction accuracy is measured simply in terms of whether $(\hat{y_{i,t+1}}- y_{i,t})\times(y_{i,t+1}-y_{i,t})>0$. In other words, if the difference between today's and yesterday's predicted value has the same sign as the difference between today's vs. yesterday's observed value, we conclude that the direction of the change was predicted accurately for that day.

\begin{figure}[!htb]
	\begin{center}
		\includegraphics[width=10cm,height=10cm]{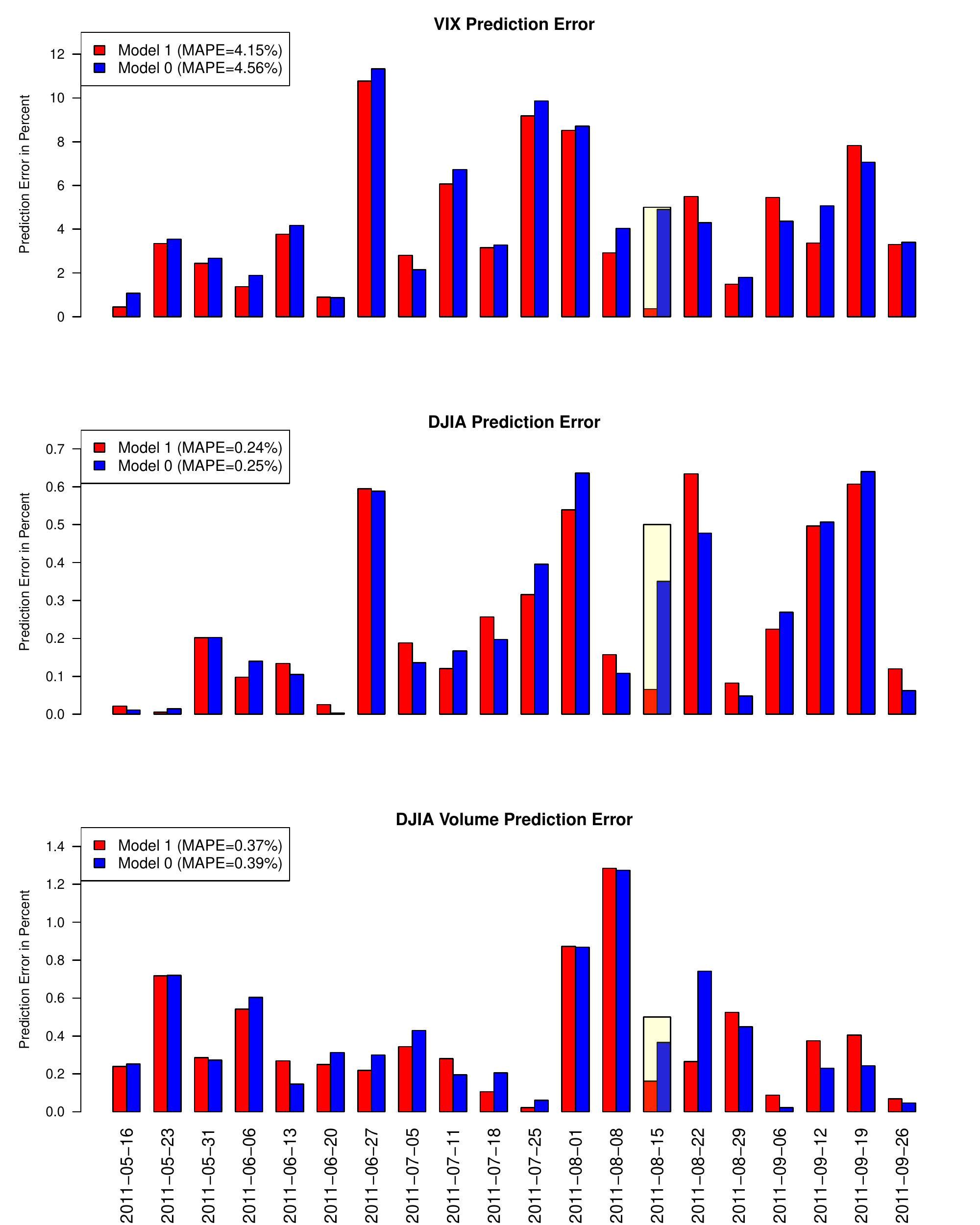} 
		\caption{\label{errorBar}Prediction Error Plot.}
	\end{center}
\end{figure}

Our search volume and financial indicator time series are available from January 2008 to September 2011. There are 196 weeks in total and we use the last 20 weeks, i.e. May 21st 2011 to October 1st 2011, as the predicting period. Each forecast uses only the information available up to the time the forecast is made. The raw data are transformed to log scale before prediction. For VIX and DJIA prediction, the lag $n$ is chosen to be 3 weeks. However, according to the Granger test analysis shown in Table \ref{grangerpGIS}, the $p-value$ is not significant for lags $>2$ weeks in the case of GIS vs. trading volume. We therefore chose $n=2$ in Eq. \ref{M0} and Eq. \ref{M1} for trading volume prediction. Fig. \ref{errorBar} shows the prediction errors for these 20 forecasting weeks. Table \ref{GISPrediction} shows the forecasting errors expressed as MAPE and direction accuracy. 

\begin{table}[!htb]
	\begin{center}
		\caption{\label{GISPrediction}Forecasting accuracy of using weekly search volumes to predict financial indicators (DJIA, Volume and VIX).}
		\begin{tabular}{cccc}
		\hline
		    		& Model   		&   MAPE     	&  Direction				 \\\hline
		\multirow{2}{*}{DJIA}  &Model~0          & 0.253   		& 0.55 \\ 
                                                      &Model~1 & 0.244 & 0.70 \\ \hline
\multirow{2}{*}{Volume}  &Model~0          & 0.386  		& 0.55 \\ 
                                                      &Model~1 & 0.366 & 0.55 \\ \hline
\multirow{2}{*}{VIX}  &Model~0          & 4.560  		& 0.55 \\ 
                                                      &Model~1 & 4.148 & 0.65 \\ \hline
		\end{tabular}
	\end{center}
	\end{table}


From these results it appears that adding search volumes (1) reduces the MAPE prediction error for VIX, DJIA and trading volumes predictions, and (2) improves the direction accuracy for DJIA and VIX forecasting, but not for trading volumes. 

Fig. \ref{errorBar} furthermore shows that during several weeks the baseline model output outperformed the advanced model. This again highlights the difficulty of financial market prediction, even using data that has been shown to have statistically significant Granger causality with the particular financial indicators. We offer the observation that on August 15th 2011 (highlighted with a yellow bar), the prediction error of the advanced model (red) dropped well below that of the baseline model (blue). In that period (August 15th -19th) the weekly VIX reached a high value of 43.05, the DJIA decreased over 450 points, and trading volumes increased significantly compared to the previous week. This is suggestive that search volumes of financial terms may be particularly useful for prediction when the market experiences high degrees of volatility, significant changes in values and high trading volumes.

\section{Twitter, Search Engine, News Media and Survey-based prediction of financial indicators}
\subsection{Correlation Analysis}
In previous sections we focused on weekly analysis due to data availability. However, our Twitter data and the Daily Sentiment Index (DSI) were recorded \emph{daily} from July 1st 2010 to September 29th 2011, for a total of 456 days. Given the availability of daily data, in this section our analysis will focus on daily time series, rather than weekly.

Again, Google Insight Search (GIS) does not provide daily volume search data. We therefore do not use GIS search volumes in our daily analyses, and instead use the Tweet volumes of financial search terms (TV-FST), as introduced in Section \ref{sec:tv-fst}.

In total, we examine four daily sentiment indicators, i.e. Twitter Investor Sentiment (TIS), Tweet Volume of Financial Search Terms (TV-FST), Negative News Sentiment  (NNS) and Daily Sentiment Index (DSI). Using the same definition as shown in Section \ref{sec:tv-fst}, the TV-FST is calculated as the average of Tweet volumes of all these financial search terms. Table \ref{corrTIS_TVFST_DSI} displays the Pearson correlation values observed between these sentiment indicators. 

\begin{table}[!htb]
	\begin{center}
		\caption{\label{corrTIS_TVFST_DSI}TIS, NNS, TV-FST, and DSI correlations.}
		\begin{tabular}{ccccc}
		\hline
		    		& TIS   		&   NNS     	&  TV-FST & DSI				 \\\hline
		TIS &   1          &    		&   &   \\ 
                   NNS & -0.237 & 1 & & \\
                   TV-FST &-0.304 & 0.225 & 1 &  \\
                   DSI  & 0.431 & -0.322 & -0.202 & 1 \\
                                                   \hline

		\end{tabular}
	\end{center}
	\end{table}

Survey data, DSI (percentage of bullish readings), has a positive correlation with TIS, but negative correlations with the other two sentiment indicators: TV-FST and NNS. TV-FST exhibits a negative correlation with DSI and TIS, but a positive correlation with NNS, which suggests that TV-FST may be a bearish/negative sentiment indicator. All listed correlations are statistically significant with $p-value < 0.01$. 

After linearly extrapolating financial indicators values missing on weekends (because of markets closing), we analyze the correlation between these sentiment indicators and financial market indexes. The results are shown in Table \ref{sentiment_Fin_corr}.

\begin{table}[!htb]
	\begin{center}
		\caption{\label{sentiment_Fin_corr}Correlations between sentiment and financial indicators.}
		\begin{tabular}{ccccc}
		\hline
		    		& DJIA   		&   Log return     	&  Volume & VIX				 \\\hline
		TIS &   -0.071          &   0.267$\star$  		& -0.127$\star$  & -0.314$\star$   \\ 
                   NNS & 0.147$\star$ & -0.147$\star$ & 0.039 & 0.237$\star$ \\
                   TV-FST &0.449$\star$ & -0.091 & 0.096 & 0.183$\star$ \\
                   DSI  & 0.277$\star$ & 0.181$\star$ & -0.341$\star$ & -0.832$\star$ \\
                                                   \hline
\multicolumn{5}{l}{($\star$ indicates $p-value < 0.01$)}

		\end{tabular}
	\end{center}
	\end{table}

We observe that TIS is positively correlated with market log returns (cf. Eq. \ref{logreturn}) and negatively correlated with VIX. DSI is positively correlated with DJIA closing values, as well as log return, but negatively correlated with trading volume and VIX. VIX reflects perceived market risk, with higher VIX values potentially indicating greater levels of investor fear. Its negative correlation with DSI and TIS may therefore indicate that the latter correspond to positive sentiment, or a lower perception of risk or fear among investors. Conversely, the positive correlation between VIX vs. NNS and TV-FST may indicate that these are indeed indicators of fear or negative sentiment.

\begin{figure}[!htb]
	\begin{center}
		\includegraphics[width=10cm,height=10cm]{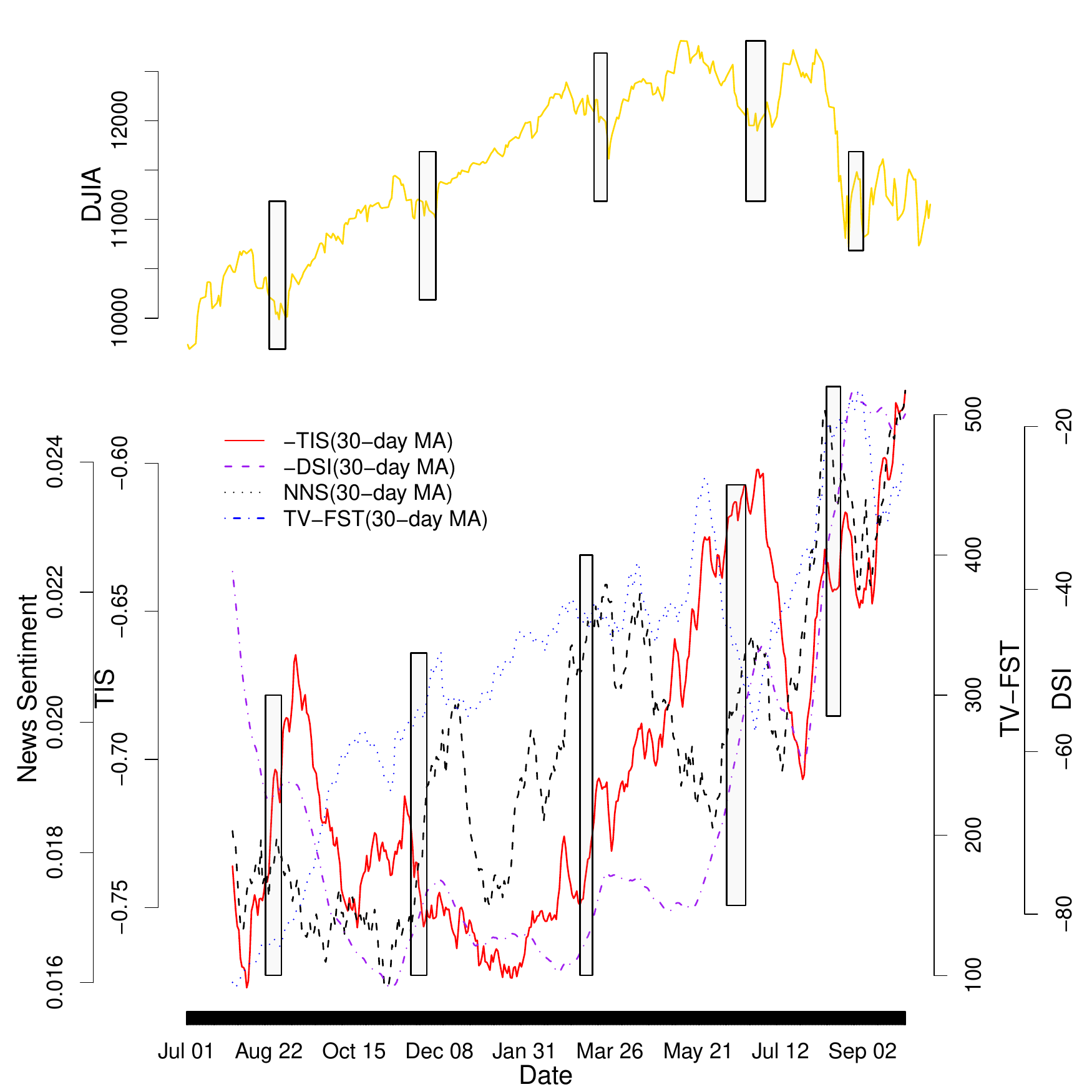} 
		\caption{\label{dailyplot}Time series of DJIA and TIS, DSI, NNS as well as TV-FST.}
	\end{center}
\end{figure}

To better view the correlation between the sentiment indicators and financial market, we plot the time series of DJIA and four sentiment indicators, in Fig. \ref{dailyplot}. 

In Fig. \ref{dailyplot}, the time series in the top panel shows the daily DJIA closing value from July 1st 2010 to September 29th 2011. The four time series in the lower panel represent TIS, DSI, NNS and TV-FST during the same time period and they are smoothed over the past 30 days. We invert the TIS and DSI to make them consistent with the directionality of the other two negative market indicators (i.e. NNS and TV-FST). As such, ``up" means negative sentiment, while ``down" indicates positive sentiment. 

We marked five time periods in the lower panel of Fig. \ref{dailyplot} with rectangle bars to indicate when DJIA prices fell in August and November 2010, and March, June and August 2011. Before DJIA prices fell in August 2010 (indicated by the first rectangle bar), it can be seen that the TIS and NNS graphs moved upwards (i.e. a rise in negative sentiment), while DSI dropped (i.e. positive). Before the second bar (November 2010), we see TIS and TV-FST trending upward. Before the fall in DJIA prices in March 2011 (third bar) we observe a clear and long-term increase of TV-FST, NNS and TIS values. TIS and TV-FST values are trending upwards before the fourth bar that marks June 2011. All four sentiment indicators trend upwards before the last bar  that makes August 2011, but the ``up" trend of DSI seem to lag the ``up" trend of NNS and TV-FST. In conclusion, though there is considerable noise in the daily data, the non-survey sentiment indicators, especially TIS and TV-FST, do show significant increases in negative sentiment that clearly precede periods of falling DJIA prices.

\subsection{Granger Causality Analysis}
The prediction of stock market returns is a matter of considerable interest. To determine whether any of our sentiment indicators are useful to predict daily DJIA log returns, we conduct a Granger causality analysis similar to Section \ref{sec:weeklyAnalysis}. According to Table \ref{sentiment_Fin_corr}, the correlation coefficient between TV-FST and log returns is statistically insignificant  ($\gamma=-0.09$). To determine which of our set of search terms are most effective to predict log returns, we conduct a correlation analysis between the search volumes of each financial term individually and log returns. Then, we select the search terms \footnote{\emph{DJIA, dow, Dow Jones, Dow Jones Industrial Average, SP500, stock(s) fall(s), stocks decline, financial market.}} whose search volumes exhibit the most significant correlations with log returns, and take the average of their time series to be the TV-FST$'$. The correlation coefficient between the resulting TV-FST$'$ and daily log returns is -0.30 with a $p$-value $<0.01$. Table \ref{grangerDaily} lists the $p$-values for a number of bi-directional Granger causality tests of log returns vs. our sentiment indicators. We find statistically significant Granger causation in both directions between log returns and TIS, NNS, and TV-FST$'$, with the exceptions of lag = 1, TV-FST$'\rightarrow$Return, and lag = 3, 5, Return$\rightarrow$TV-FST$'$. No statistically significant Granger causation was observed between DSI and log returns. These results indicate that sentiment indicators extracted from Twitter (TIS and TV-FST$'$) and News headlines (NNS) are predictive to the DJIA log return, but DSI is not predictive. 

\begin{table*}
	\begin{center}
		\caption{\label{grangerDaily}Statistical significance (p-values) of  Granger causality analysis  between daily log return and TIS, NNS, DSI and TV-FST$'$.}
		\begin{tabular}{cccccc}
		\hline
		    		& 1   		&   2     	&  3	 & 4 & 5 			 \\\hline
		TIS$\rightarrow$Return          & \textless{0.001}$\star\star\star$    		& 0.0086$\star\star\star$        	& 0.035$\star\star$ & 0.028$\star\star$ & 0.021$\star\star$
    					\\
		Return$\rightarrow$GIS      	&   \textless{0.001}$\star\star\star$     	& \textless{0.001}$\star\star\star$  		&   \textless{0.001}$\star\star\star$ & \textless{0.001}$\star\star\star$ & \textless{0.001}$\star\star\star$     						\\
		  NNS$\rightarrow$Return      &  0.017$\star\star$    		& 0.030$\star\star$      		& 0.011$\star\star$  &0.005$\star\star\star$  	 & 0.004$\star\star\star$
\\
		Return$\rightarrow$NNS           & 0.014$\star\star$     	& 0.013$\star\star$   		& 0.031$\star\star$ & 0.030$\star\star$ & 0.055$\star\star$  		 			\\
		DSI$\rightarrow$Return     &  0.523     	& 0.138   		& 0.203 & 0.308 & 0.377 \\  					
		Return$\rightarrow$DSI        & 0.267     	& 0.174  		& 0.647 & 0.377 & 0.059 $\star$   		\\
TV-FST$'\rightarrow$Return & 0.413 & \textless{0.001}$\star\star\star$ & \textless{0.001}$\star\star\star$ & \textless{0.001}$\star\star\star$ & \textless{0.001}$\star\star\star$ \\
Return$\rightarrow$TV-FST$'$& 0.0025$\star\star\star$ & 0.019$\star\star$ & 0.151  & 0.071$\star$ & 0.140 
\\\hline
\multicolumn{4}{l}{ ($p-value<0.01$: $\star\star\star$, $p-value< 0.05$: $\star\star$, $p-value<0.1$: $\star$)}
		\end{tabular}
	\end{center}
	\end{table*}

\subsection{Multiple Regression Analysis}

In this section, we conduct a multiple regression for daily log returns obtained according to Eq. \ref{logreturn}. The regression inputs are our four sentiment indicators and the past financial values of log return. As an additional control, we include VIX, since it is a well-accepted predictor for market return. The multiple regression model is shown in Eq. \ref{multipleRegressionEq}, where $n = 7$ days, and $Y$ represents the daily log return. In order to maintain a common scale, we normalized all data to standard scores. 

\begin{equation}
\begin{split}
Y_{t}=\alpha+\sum_{i}^n\beta_{i}Y_{t-i} +\sum_{i}^n\chi_{i}TIS_{t-i}+\sum_{i}^n\delta_{i}NNS_{t-i}+ \\ \sum_{i}^n\phi_{i}TV-FST'_{t-i}+\sum_{i}^n\gamma_{i}DSI_{t-i}+\sum_{i}^n\eta_{i}VIX_{t-i}+\epsilon_{t}
\label{multipleRegressionEq}
\end{split}
\end{equation}  

Table \ref{multiple_regression} provides the summary statistics of the multiple regression. Compared with the baseline model, the adjusted $R^2$ improves from 0.092 to 0.200. This means that an additional 11\% of the variation in log returns is accounted for by adding these sentiment indicators. After controlling for all other variables, we find that DSI is not a statistically significant predictor. The two sentiment indicators extracted from Twitter, i.e. TIS and TV-FST$'$, are however very significant predictors at a lag of 1 to 2 days. Here, we observe a reversal effect, namely that daily log returns are positively associated with TIS and TV-FST$'$ on the previous day, but negatively correlated with those on the lag of 2 days. VIX values at lags of 2 days are highly statistically significant predictors of log return. NNS is also a statistically significant predictor at lags ranging from 1 or 4 days, but with much less lower coefficients, e.g. at a lag$=1$ we find that the $p-value=0.08$, and the coefficient is -0.087, which means we expect to see a log return decrease of only 0.087 standard deviations for each one standard deviation increase of NNS. 

\begin{table*}
	\begin{center}
		\caption{\label{multiple_regression}Summary statistics of multiple regression.}
		\begin{tabular}{c|cc|cc|cc|cc|cc|cc}
		\hline
		    		Lag & \multicolumn{2}{c|}{Return} & \multicolumn{2}{c|}{TIS} & \multicolumn{2}{c|}{NNS} & \multicolumn{2}{c|}{DSI} & \multicolumn{2}{c|}{VIX} & \multicolumn{2}{c}{TV-FST$'$}  \\ \hline
                                           & Coeff. & $p$-value & Coeff. & $p$-value & Coeff. & $p$-value & Coeff. & $p$-value & Coeff. & $p$-value & Coeff. & $p$-value  \\ \hline
1 & 0.282 & 0.004$\star\star\star$ & 0.170 & 0.008$\star\star\star$ &-0.087&0.080$\star$ & 0.389 & 0.385 &0.494 & 0.182 & 0.235 & 0.0007$\star\star\star$ \\
2 & -0.139 & 0.175 & -0.164 & 0.018$\star\star$ &-0.066 & 0.191 &-0.239 &0.709 & -1.161 & 0.029$\star\star$ & -0.324 &3.73e-05$\star\star\star$ \\
3 & 0.0006 & 0.995 & 0.069 & 0.316 & 0.048 & 0.349 & -0.265 & 0.678 & 0.641 & 0.235 & 0.059 & 0.486 \\
4 & -0.115 & 0.275 & -0.088 & 0.208 & -0.097 & 0.058 $\star$&0.730 & 0.251 & 0.384 & 0.479 & -0.059 & 0.490 \\
5 & -0.0212 & 0.837 & 0.152 & 0.031$\star\star$  & -0.017 & 0.740 & 0.177 &0.780 &0.502 &0.351 & 0.171 &0.045$\star\star$ \\
6 & 0.071 & 0.472 & -0.132 & 0.062 $\star$ & 0.057 & 0.257 & -0.514 &0.405 & -0.490 & 0.341 & -0.096 & 0.261 \\
7 & -0.117 & 0.040$\star\star$ & 0.005 & 0.935 & 0.008 & 0.874 & -0.115 & 0.789 & -0.204 & 0.559 & -0.024 & 0.743 \\   
                                                   \hline
\multicolumn{12}{c}{($p-value<0.01$: $\star\star\star$, $p-value<0.05$: $\star\star$, $p-value<0.1$: $\star$)} \\
\multicolumn{12}{c}{Residual standard error: 0.893 on 406 degrees of freedom.} \\
\multicolumn{12}{c}{Multiple $R$-squared: 0.2841, Adjusted $R$-squared: 0.200 (baseline model (cf. Equation \ref{M0}, $Y$ is daily log return here): Adjusted $R$-squared 0.092).} \\
\multicolumn{12}{c}{$F$-statistic: 3.67 on 42 and 406 DF, $p$-value: 5.523e-12} \\
\hline
		\end{tabular}
	\end{center}
	\end{table*}

\subsection{Forecasting analysis}

To further test the hypothesis that adding sentiment indicators can help predict financial indicators such as the DJIA, trading volumes, and VIX, we conduct a 1-step forecasting test over 30 days, i.e. from August 31st 2011 to September 29th 2011. As with the weekly prediction in Section \ref{sec:weeklyPred}, the baseline model is based on its own historical financial values (cf. Model 0 in Eq. \ref{M0}) whereas the advanced model (cf. Model 1 in Eq.\ref{M1}) adds the historical values of the sentiment indicators TIS, NNS, TV-FST and DSI. Here we assume $n=7$. The forecasting accuracy is measured in terms of the Mean Absolute Percentage Error (MAPE) and the direction accuracy. Results are shown in Table \ref{dailyPrediction}.

\begin{table}[!htb]
	\begin{center}
		\caption{\label{dailyPrediction}Forecasting accuracy of using TIS, NNS, TV-FST and DSI to predict financial indicators (DJIA, trading volume and VIX).}
		\begin{tabular}{cccc}
		\hline
		    		& Model   		&   MAPE     	&  Direction				 \\\hline
		\multirow{2}{*}{DJIA}  &Model~0          & 1.00   		& 0.5 \\ 
                                                      &Model~1 & 0.97 & 0.63 \\ \hline
\multirow{2}{*}{Volume}  &Model~0          & 7.24  		& 0.47 \\ 
                                                      &Model~1 & 7.56 & 0.60 \\ \hline
\multirow{2}{*}{VIX}  &Model~0          & 4.00  		& 0.6 \\ 
                                                      &Model~1 & 3.88 & 0.67 \\ \hline
		\end{tabular}
	\end{center}
	\end{table}

 We find improvements in the direction accuracy and MAPE of the forecasting accuracy for DJIA, VIX and volume prediction, with the exception of the MAPE for volume prediction. However, the improvement is not highly significant. The extremely high volatility in the financial markets during our training and testing periods, especially in August and September 2011, may account for this. In addition, we used relatively simple linear models in this paper that may not be suited to model the complex interactions of factors involved in shaping financial market values. Further research will need to focus on the development of more accurate and more advanced linear/non-linear prediction models.

\section{Conclusion}
Behavioral finance challenges the Efficient Market Hypothesis by emphasizing the important role that human emotion, sentiment and mood play in financial decision-making. Thus places the accurate measurement of sentiment and mood at the heart of a discussion over how to best model and predict the behavior of the financial markets. Previous research in this domain has relied mainly on surveys or news analysis to obtain investor sentiment. Research has recently started to leverage very large-scale web data, including search engine and social media data, to assess public as well as investor sentiment. However, most existing work adopts only a single data source (survey, social media or search engine data) as a proxy to public and investor sentiment, and then uses it to computer a particular financial index. To the best of our knowledge, no work has been done to perform a detailed survey of a variety of different classes of mood indicators extracted from a variety of classes of data sources. Studying the relations between different mood indicators and their predictive relationships to different financial indexes is necessary to unravel the causal relations sentiment and mood relate to the financial markets, and thus crucial in improve financial forecasting models. Our paper is a first, preliminary contribution of such a comparison to the rapidly emerging domain of \emph{computational behavioral finance}. 

In this paper, we collect six sentiment indicators from investor sentiment surveys (II and DSI), social media (Twitter), news media services, as well as search engine (Google). Those include DSI bullish percentage, Investor Intelligence (II), Twitter Investor Sentiment (TIS), Tweet volumes of financial search terms (TV-FST), Negative News Sentiment (NNS) and Google search volumes of financial search terms (GIS). 

First, in a weekly analysis, we find a significant correlation between weekly GIS of financial terms with DJIA closing values, trading volume, and VIX values. Granger causality tests confirm that GIS is indeed predictive of financial indicators, but surveys of investor sentiment (i.e. II) are not. Weekly forecasting accuracy was improved by adding GIS search volumes of financial terms, notably so when the DJIA was trending downward and the VIX was indicating high volatility, such as in August 2011.

Second, in our daily analysis, all mood indicators exhibited a significant correlation with log returns and VIX values. Controlling for other mood indicators including the VIX, we find that TIS and the TV-FST values of the previous 1-2 days are very statistically significant predictors of daily market returns, while DSI is not. NNS is also found to be a statistically significant predictor, however, compared to the TIS and TV-FST, we find less significant predictability of log return. This finding indicates that the predictive power of Twitter's two sentiment indicators outperformed survey sentiment as well as news media analysis. Moreover, we found that before the highly downward movement of DJIA in the end of July and August 2011, Tweet volumes of financial terms started to increase several weeks earlier than Google volumes did. This indicates a potential efficiency gain of Twitter over GIS. 

Studying the predictive power of the online web data is still in its infancy. The various correlations and limitations of these different data sources, different sentiment measures, and its general prediction applicability to different domains remain unclear. Our work is the first attempt to extract a range of sentiment indicators from several popular data sources (Twitter, search engine and news) and use various sentiment indicators to predict different financial indexes (DJIA, trading volumes, VIX and gold) in both daily and weekly scale. Continued research is needed to deepen our understanding of why and how these different mood indicators relate to and predict socio-economic phenomena such as the financial markets. 

\section*{Acknowledgements}
We would like to express our gratitude to trade-futures and Investor Intelligence for making their sentiment data available to our research program. We also thank the Socionomics Institute for their kind advice on a range of research questions, and their assistance in obtaining research data that was crucial in conducting the research presented in this paper.

\bibliographystyle{abbrv}
\bibliography{mao.bib}

\begin{biography}[{\includegraphics[width=1in,height=1.25in,clip,keepaspectratio]{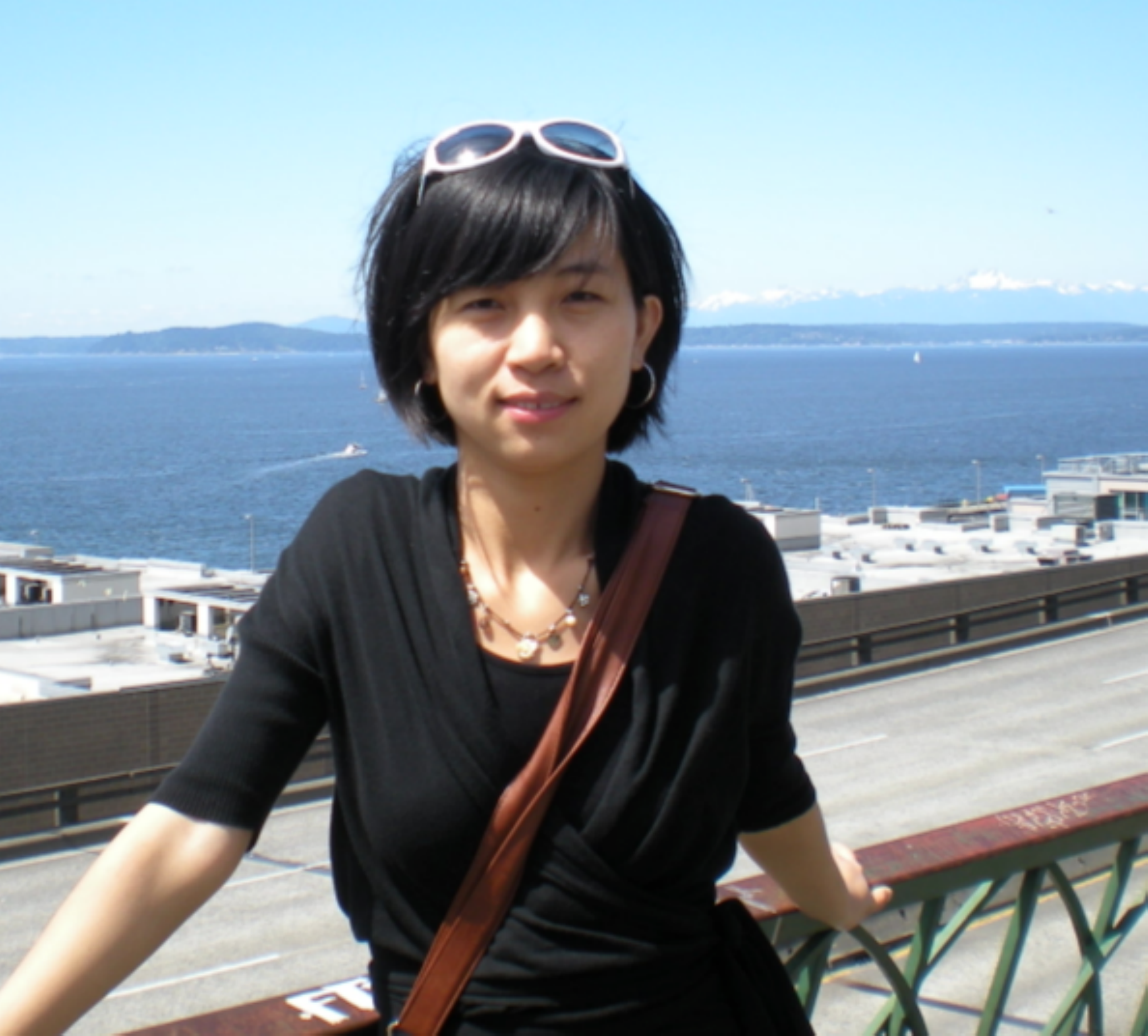}}]{Huina Mao} 
School of Informatics and Computing, Indiana Unviersity-Bloomington, United States.
\emph{Email: huinmao@indiana.edu}
\end{biography}

\begin{biography}[{\includegraphics[width=1in,height=1.25in,clip,keepaspectratio]{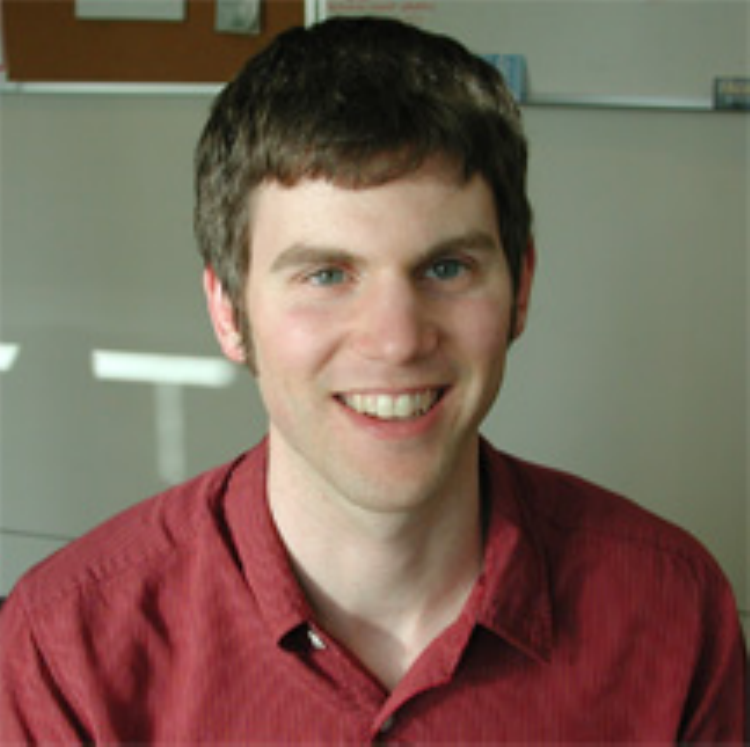}}]{Scott Counts} 
Microsoft Research, Redmond, Washington, United States.\newline
\emph{Email: counts@microsoft.com}
\end{biography}

\begin{biography}[{\includegraphics[width=1in,height=1.25in,clip,keepaspectratio]{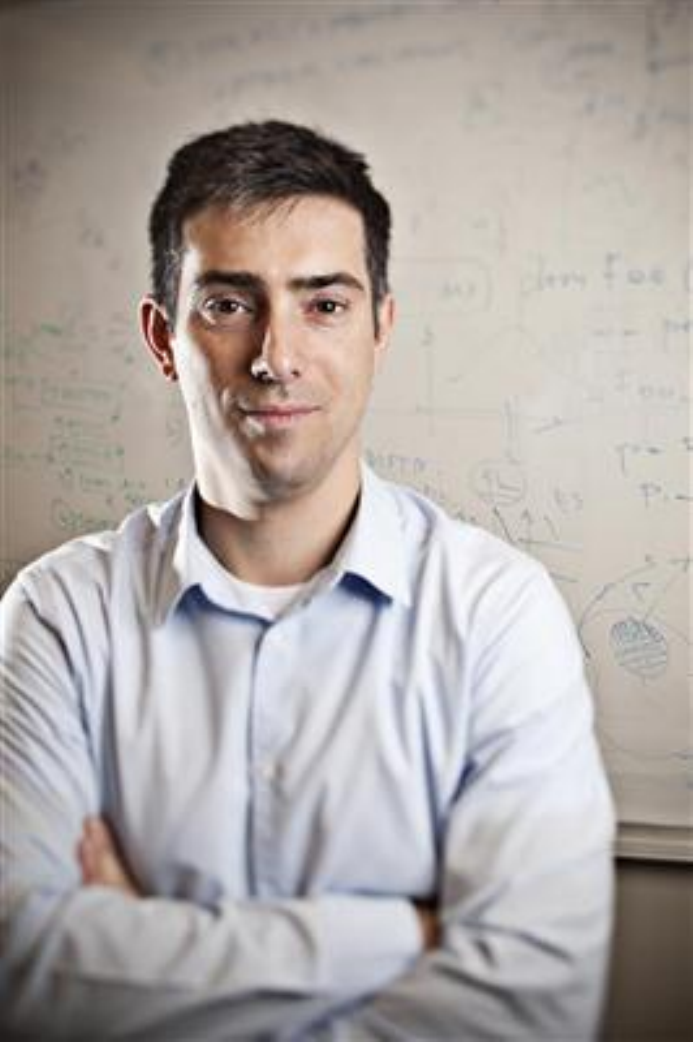}}]{Johan Bollen} 
School of Informatics and Computing, Indiana Unviersity-Bloomington, United States. \emph{Email:jbollen@indiana.edu}
\end{biography}
\end{document}